\documentclass[10pt,twocolumn]{article}

\usepackage[margin=0.75in]{geometry}
\usepackage{graphicx}
\usepackage{booktabs}
\usepackage{amsfonts}
\usepackage{amsmath}
\usepackage[hyphens]{url}
\usepackage{hyperref}
\usepackage{authblk}

\usepackage{balance} 

\title{Deep Learning Based Auction Design for Selling Agricultural Produce through Farmer Collectives to Maximize Nash Social Welfare}

\author[1]{Mayank Ratan Bhardwaj}
\author[2]{Vishisht Srihari Rao}
\author[2]{Bazil Ahmed}
\author[2]{Kartik Sagar}
\author[2]{Y. Narahari}

\affil[1]{Plaksha University}
\affil[2]{Indian Institute of Science}

\date{}

\begin{document}

\maketitle

\begin{abstract}
This paper is motivated by the need to design a robust market mechanism to benefit farmers (producers of agricultural produce) as well as buyers of agricultural produce (consumers). Our proposal is a volume discount auction with a Farmer Collective (FC) as the selling agent and high volume or retail consumers as buying agents. An FC is a cooperative of farmers coming together to harness the power of aggregation and economies of scale. Our auction mechanism seeks to satisfy fundamental properties such as incentive compatibility and individual rationality, and an extremely relevant property for the agriculture setting, namely, Nash social welfare maximization. Besides satisfying these  properties, our proposed auction mechanism also ensures that certain practical business constraints are met. Since an auction satisfying all of these properties exactly is a theoretical impossibility, we invoke the idea of designing deep learning networks that learn such an auction with minimal violation of the desired properties. The proposed auction, which we call VDA-SAP (Volume Discount Auction for Selling Agricultural Produce), is superior in many ways to the classical VCG (Vickrey-Clarke-Groves) mechanism in terms of richness of properties satisfied and further outperforms other baseline auctions as well. We demonstrate our results for a realistic setting of an FC selling perishable vegetables to potential buyers.
\end{abstract}


\section{Introduction}\label{sec:Intro}
Small and marginal farmers generally own up to 5 (approx. 2 hectares) of land and grow 1 or 2 crops at a time. In terms of selling the harvested produce, they may not always be able to directly reach consumers and even if they do, they may not be able to negotiate a competitive price.  Consumers can be classified into three categories based on the volumes they purchase. 
The first category of consumers is made up of large chains of retail and grocery stores who buy large volumes of the produce and redistribute it through several outlets or online platforms. 
The second category of consumers comprises those who own standalone retail stores and generally sell to individual consumers in a certain locality. The third category consists of the individual consumers  who generally purchase a smaller quantity that suffices to satisfy their household needs. 
In reality, it is not practical for an individual farmer to be matched directly with consumers as several logistical issues arise. Some examples of these are a farmer not being able to reliably satisfy all the needs of the matched consumers, large companies looking to purchase large volumes of a particular grade of produce, etc. For individual farmers to reach individual consumers, several other issues may arise such as transportation logistics and the preference or urge to sell all of the harvested crop. It is important to find a way of selling the produce that maximizes the social welfare (combined utilities of farmers and consumers). 

For these reasons, it is essential for an organisation or official intermediary to step in and solve the problem of selling produce to all potential consumers in a way that is competitive and attractive for the farmers as well as the consumers. 
This is one of the primary objectives of Farmer Collectives (FCs), or Farmer Producer Organisations (FPOs), 
which have been set up in different parts of the world.

An FPO or FC can be thought of as an entity that represents a large group of small and marginal farmers (generally between 100 to a few thousand farmers) from a certain geographical region.  Grouping of small farmers into FPOs or 
FCs can greatly support small farmers in modern agricultural markets, through functions that small farmers cannot perform individually, such as procuring high-quality inputs closer to villages, aggregation and marketing of produce, facilitating access to infrastructure and technology, and providing training~\cite{IDINSIGHT19}. From now on, we use the acronym FC (rather than FPO) to maintain uniformity.

Reports suggest there exist  more than 1500 FCs in Brazil, more than 7500 FCs in Germany, more than 63000 FCs in India, and more than 1700 FCs in USA. 
In fact, \cite{EarthObservingSystem22} mentions a staggering 1.2 million FCs across the globe today. Harnessing the potential of FCs through collective action to make marketing of harvested produce more attractive for farmers as well as consumers has high potential for visible impact.


Since scientifically designed market/auction mechanisms can promote honest behavior and healthy competition among buyers and sellers~\cite{MILGROM89}, our proposal 
is to develop a suitable auction mechanism for selling harvested produce through intermediation by FCs. During our interactions with some FCs, the FC administrations have mentioned that such auctions would be very useful for perishable crops such as tomatoes, onions, brinjal, chili peppers, seasonal fruits, etc.

This paper proposes an auction mechanism that can help small and marginal farmers sell their harvested produce while obtaining a competitive price that also remains attractive to the consumers. Currently, auction mechanisms are not very popular for selling harvested produce; however, an intermediary such as an FC will make auctions a viable option. 

The auction mechanism proposed is a volume discount auction where bids are invited from consumers, possibly seeking discounts based on volumes.   
We call our mechanism VDA-SAP (Volume Discount Auction for Selling Agricultural Produce).
VDA-SAP seeks to satisfy the following properties: incentive compatibility, individual rationality, Nash social welfare (NSW) maximization, fairness, and business constraints. Satisfying all of these properties exactly, according to mechanism design theory~\cite{KRISHNA09,NARAHARI14} 
is a theoretical impossibility. Hence, we innovatively adapt a deep learning based approach (originally proposed by \cite{DUTTING15, DUTTING19, DUTTING21}) that minimizes a loss function that captures the violation from the desired properties. We show that our mechanism outperforms a standard VCG (Vickrey-Clarke-Groves) mechanism even while satisfying several additional properties such as Nash social welfare maximization and several pertinent business constraints.

\section{Relevant Work and Contributions}
Several  countries have experimented with online platforms for selling harvested  produce so as to benefit farmers and consumers. Examples include exchange platforms in Ethiopia~\cite{ECONOMIST17}, Kenya~\cite{MAINA23}, and Uganda~\cite{NEWMAN18}, the eNational Agriculture Market (eNAM) in India~\cite{ENAM}, and the Unified Market Platform (UMP) in the state of Karnataka, India~\cite{LEVI22}. The mechanisms used in these platforms sweep a wide spectrum:  warehouse-based negotiation, ascending auctions, first-price sealed-bid auctions, etc.~\cite{LEVI22}.  It is not clear which mechanism is most beneficial for farmers~\cite{STRZKEBICKI15}.
Empirical evidence shows that the mechanisms may or may not be beneficial. As a case in point, the Ethiopia Commodity Exchange is very popular for coffee and sesame seeds and not popular for any other crops~\cite{ECONOMIST17}.
In India, for example, only 14\% of all farmers have registered on eNAM, and seemingly, over half of these registered farmers  have not benefited from the platform~\cite{JADHAV19}.

Kudu is an agricultural marketplace in Uganda~\cite{NEWMAN18} in which farmers and traders use their mobile devices to place bids (requests to buy) and asks (requests to sell) using a centralized nationwide database.  Kudu identifies profitable trades, which are proposed to the participants.   
An automatic matching algorithm that uses a maximum weight matching algorithm in a bipartite graph takes as input a set of bids and asks, and algorithmically proposes trades.  

~\cite{VISWANADHAM12} provides a good overview of the working of the Indian Agricultural markets, which are  called mandis. It proposes that mandis should be transformed into electronic exchanges and present a mixed integer programming model that the electronic exchange needs to solve in an iterative way to optimally match buyers with sellers. It also presents a stylized case study to illustrate the functioning of such an electronic exchange.

~\cite{DEVI15} proposes a matching algorithm that innovatively uses the Gale Shapley algorithm~\cite{GALE62}. The results obtained using this approach outperform the results obtained using an English auction based method. It is found that the proposed method produces stable matching, which is preference-strategy proof and it also reduces the need for number of rounds of allocation.

\cite{LEVI22} introduces a behavior-centric, field-based, and data-driven methodology to propose and design auction mechanisms that enhance the revenue of farmers in online agricultural platforms. It proposes and implements a new two-stage auction for the agri platform for the Karnataka State in India for a major lentils market. The implementation saw the participation of more than 10,000 small and marginal  farmers in the market in three months time. 

~\cite{ZHANG21} and~\cite{SANKAR24} provide a brief overview of some common deep learning architectures used in auction design. In particular, the latter discuss deep learning based procurement auctions in the agricultural domain.

\subsection{Contributions and Novelty}
In this paper, we demonstrate that an FC can act as a convenient intermediary between the farmers and consumers with the help of a sound protocol for selling harvested produce. 
Our idea is simple: An FC will aggregate (virtually, rather than, physically) the harvested produce from its farmer members (who are willing) and sell the aggregated commodities to potential consumers using a mechanism such as auctions. In particular, the mechanism we propose is a volume discount  auction that we call VDA-SAP (Volume Discount Auction for Selling Agricultural Produce).

A volume discount auction mechanism has earlier been used in ~\cite{BHARDWAJ23E}, but in the context of agricultural input procurement. The situation considered there is, in a manner, the reverse of the situation considered in this work.  In~\cite{BHARDWAJ23E}, the input requirements of farmers (for example, seeds, fertilizers, or pesticides) are aggregated by the FC and the aggregated requirements are bulk procured in a cost-effective way from suppliers through a volume discount auction that is a reverse auction. The volume discount reverse auction proposed there seeks to satisfy incentive compatibility, individual rationality, cost minimization, fairness, and business constraints, with a slight compromise in social welfare maximization. The auction is designed using a deep learning based approach. 

The proposed VDA-SAP mechanism  is a volume discount (forward) auction that uses a deep learning approach with the novelty that instead of cost minimization or revenue maximization, we seek Nash social welfare (NSW) maximization, in addition to other desirable properties. NSW (see~\cite{CARAGIANNIS19}) for a mechanism is the geometric mean of the utilities of the agents and satisfies several nice properties that are extremely pertinent to agricultural context: (a) NSW maximization balances the requirements of the farmers as well as the consumers while not compromising on social welfare (b) NSW maximization also has very appealing fairness properties. This ensures that VDA-SAP favors farmers as well as consumers and therefore becomes sustainable.  


In addition to NSW maximization, the proposed mechanism seeks to satisfy (a) dominant strategy incentive compatibility (b) ex-post individual rationality (c) fairness, and  (d) practical business constraints. Since no mechanism can theoretically satisfy all the above properties exactly, we invoke a deep learning approach for auction design, a technique that is now becoming popular in the literature (see for example \cite{BHARDWAJ23E, ZHANG21, DUTTING21, DUTTING19, DUTTING15}). We show, using a realistic setting of selling a perishable vegetable that our proposed mechanism outperforms the canonical VCG auction as well as several other baseline auctions that we progressively introduce.
In our view, the proposed mechanism, when properly implemented on the ground, will significantly benefit producers and consumers of perishable agricultural produce. 

The rest of the paper is organized as follows. Section~\ref{sec:Prosper} describes the proposed VDA-SAP in terms of the structure of the bids and the desirable properties we seek to satisfy. Section~\ref{sec:DLApproach} sets up the deep learning based approach for learning different variants of the auction, VDA-SAP. Section~\ref{sec:Exp} presents the experimental results for a realistic scenario of selling 1000 kg of chili pepper using a VCG auction and several variants of VDA-SAP.

\section{VDA-SAP Mechanism}\label{sec:Prosper}
During our initial interactions with a few FCs, we realized that the proposed auction, VDA-SAP would work best for perishables such as vegetables. Hence, to explain VDA-SAP, we provide the example of an FC whose farmers largely grow vegetables that are sold by the FC. 
For each vegetable that is grown by the farmers of the collective, a separate auction is conducted. Let us, for the sake of this example, consider the sale of tomato. 
The consumers in the auction belong to the first two categories of consumers mentioned in Section~\ref{sec:Intro}, i.e., large chains of retail stores and standalone retail stores.
We do not include small individual consumers in the auction marketplace since an auction is not required for the small individual consumers. Instead, a simple e-commerce platform would be most convenient there.

\begin{figure}[h]
        \includegraphics[width=\columnwidth]{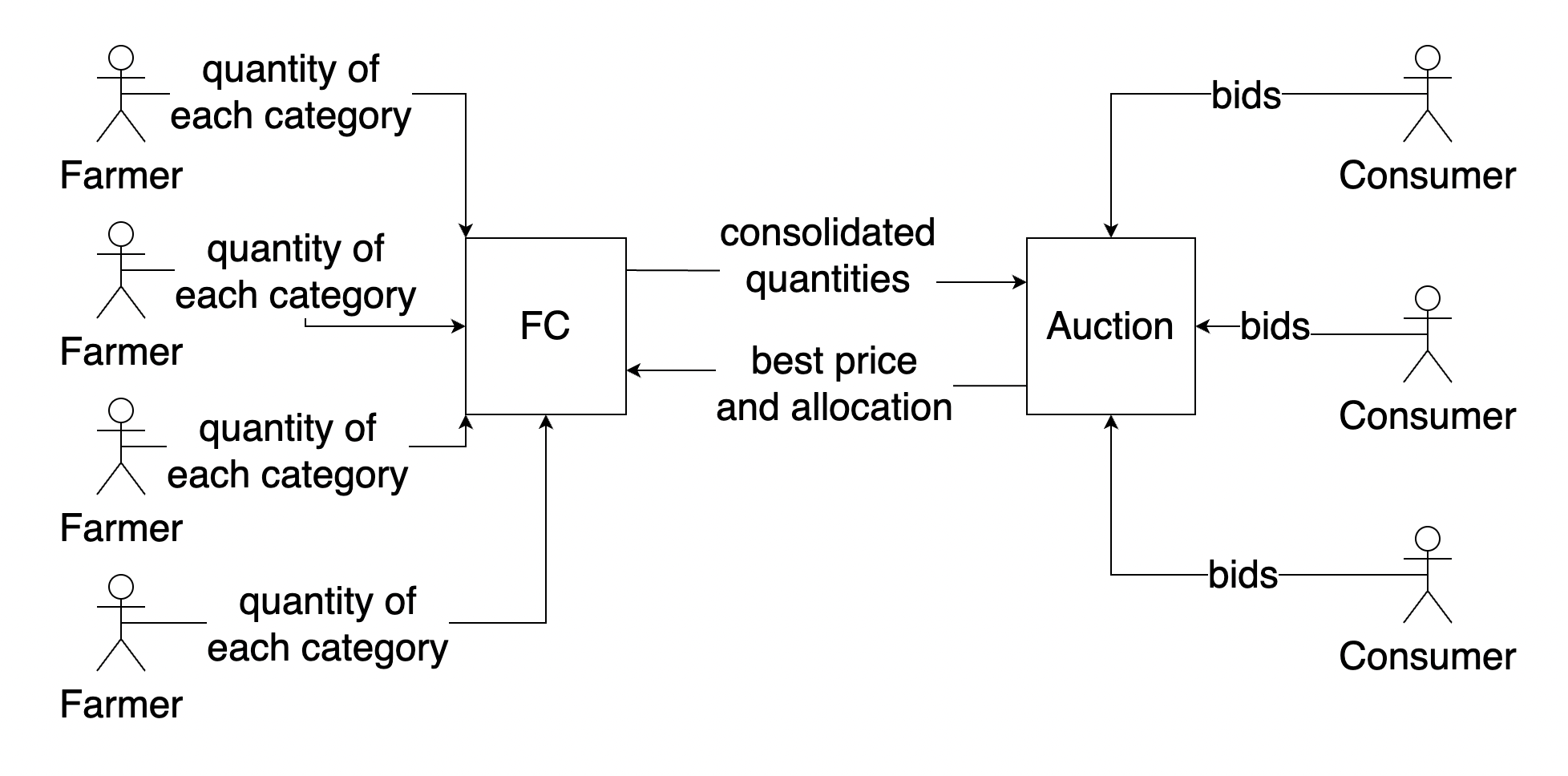}
        \caption{VDA-SAP for maximizing Nash social welfare}
        \label{fig:FPOauction}
    \end{figure}

The setup is as follows. Individual farmers approach the FC with the produce that they have grown. The FC records the quantities brought by the individual farmers and accumulates all the produce. 
Following this accumulation, the FC announces, to all potential consumers, the total volume available for sale. Each consumer is asked to place a bid, which includes the quantity of produce required and the price per unit that the consumer is willing to pay for that quantity.
Large scale consumers may usually obtain the daily market values based on current supply and demand in the market and submit their bids based on these values. Volume discounts may also be sought as part of the bids. Figure~\ref{fig:FPOauction} captures the workflow of VDA-SAP.

\subsection{VDA-SAP: Two Example Bids}
As an example of a volume discount bid, let us assume that 100 kg of tomato is available to be sold by the FC. A particular consumer may bid \$ 5.5/kg for the first 60 kg and \$ 5/kg for any further tomatoes bought up to 100 kg.  We call the first 60 units as the first lot and the next 40 units as the second lot. If allotted 75 kg of tomatoes, this consumer is ready to pay \$ 5.5 $\times$ 60 $+$ \$~5 $\times$ 15 = \$ 405. 

Different consumers may use different bidding languages. For example, a consumer planning to purchase 100 kg of tomato may bid \$ 5.25/kg for up to 50 kg and \$ 5/kg if the quantity is more than 50 kg. This means that if 75 kg is allotted to 
the consumer, then the consumer is ready to pay \$ 5/kg $\times$ 75 = \$ 375.
This is an example of a flat discount. It is worth noting that all the bids can be converted to the same bidding language before running the auction. 

There are several factors to be considered while determining who is allocated how much of the produce, and at what price. These will be elaborated in the rest of the paper.

\subsection{VDA-SAP: Desirable Properties}

We mention below the most desirable properties. Our analysis is based on our familiarity with and study of FCs and the dynamics of selling agricultural produce. 

\subsubsection*{Incentive Compatibility  (IC)} 
Incentive compatibility ensures truthful bidding by the consumers and is a fundamental requirement of any auction mechanism. The most powerful version of IC is dominant  strategy incentive compatibility (DSIC). DSIC ensures that, irrespective of the bids of the other consumers, it is in the best interest of each consumer to bid their true value.

\subsubsection*{Individual Rationality (IR)}
Individual rationality ensures that the FC and consumers obtain non-negative utility by participating in the auction. The most powerful version of IR is ex-post IR, which implies that the utility to each participating player (all the consumers as well as the FC) will be non-negative irrespective of the actions of the other players.

\subsubsection*{Social Welfare Maximization (SWM)}
SWM aims to maximize the sum of utilities of the FC and all the consumers. 
The utility of the FC captures the joint utility of all the farmers whose produce is aggregated by the FC. 

\subsubsection*{Nash Social Welfare (NSW) Maximization}
Although social welfare maximization satisfies the utilitarian angle, it allows a particular player's utility to take a hit if another player is receiving a commensurate or greater utility boost. A more egalitarian approach would be to maximize the product of utilities of all the participants in the auction. In this paper, we focus on NSW maximization rather than social welfare maximization since NSW maximization is known to benefit both the seller (FC on behalf of farmers) and the consumers as it approximates the Maximin guarantee~\cite{CARAGIANNIS19}. \cite{CARAGIANNIS19}, \cite{YUEN23} and other recent literature have highlighted the virtues of NSW maximization. 

\subsubsection*{Revenue Maximization for Farmers (FOPT)}
It is desirable that the expected total revenue generated for the farmers is maximized. FOPT makes the auction farmer-friendly.

\subsubsection*{Cost Minimization for Consumers (COPT)}
For running multiple successful iterations of the mechanism, it is imperative that the expected total cost incurred by the consumers is minimized. COPT implies a consumer-friendly auction.

\subsubsection*{Business Constraints (BUS)}
It would help the FC run the auctions effectively over a longer time duration if certain business rules were implemented. 
Business constraints refer to constraints such as having a minimum number of winning consumers (to avoid monopoly by a single or small number of consumers),  a maximum number of winning consumers (to minimize logistics costs), a maximum fraction of business to be awarded to any consumer, etc. 

\subsubsection*{Fairness (FAIR)}
Fairness implies that the winning consumers are chosen in a fair way. An index of fairness would be envy-freeness -- no consumer can increase their utility by adopting another consumer's outcome. If envy-freeness is not achievable, the next best option is envy minimization (which is what we pursue in this paper).
It should be noted that Nash Social Welfare and Business Constraints also, in their own respective ways, promote fairness to a certain extent. However, envy minimization is a popular, exclusive  measure of fairness.

\subsubsection*{Properties to Satisfy: Possibilities and Impossibilities}
Ideally, we would like VDA-SAP to satisfy DSIC, IR, SWM, NSW maximization, FAIR, and BUS. The classical Vickrey-Clarke-Groves (VCG) mechanisms 
satisfy DSIC, IR, and SWM but may not  satisfy FAIR, NSW maximization, and BUS. 
Satisfying the set of properties (DSIC, IR, SWM, NSW maximization, FAIR, BUS) is clearly a tall order. Mechanism design theory~\cite{KRISHNA09,NARAHARI14} 
is replete with impossibility theorems which make it clear that all these properties cannot be satisfied simultaneously. We therefore take the approach of satisfying as many properties as possible while minimizing the violation of other properties. This leads to several variants of VDA-SAP, which are described later in the paper. We invoke a deep learning based approach to learn the allocation rules and payment rules of these auctions.

The properties FOPT and COPT make the auction farmer friendly or consumer friendly, respectively. They have been included for the sake of completeness and for comparison purposes.

\section{A Deep Learning Approach for VDA-SAP} \label{sec:DLApproach} 
We propose a deep learning based approach to auction design that is an extension of \cite{FENG18,DUTTING21,BHARDWAJ23E}. It is to be noted that all these mechanisms guarantee IR. \cite{DUTTING21} maximizes the revenue of the seller in the auction while minimizing the violation of IC. 
The architecture we use for 
VDA-SAP is similar to the architectures used by \cite{DUTTING21} and \cite{BHARDWAJ23E}. However, it is to be noted that these architectures cannot be used as is, because of the following reasons.

In VDA-SAP we are dealing with the selling  of homogeneous units with volume discounts, whereas \cite{DUTTING21} considers auctions with additive (or unit-demand) valuations.  Our volume discount auction setting is not additive and is not unit-demand either. To see this, observe that the value  of $2x$ units with volume discounts is not the same as twice the value of $x$ units. Additionally, the consumers in our setting wish to buy any number of units, unlike the unit-demand buyers considered by~\cite{DUTTING21}. Auctions with additive valuations and unit-demand valuations make it possible to use allocation networks whose outputs are simply (stochastic) allocation matrices. Since we cannot do this in our case, we produce an allocation tuple as output - with each element in the tuple being the allocation for the corresponding consumer. This complicates the computation of the payments and makes theoretical analysis non-trivial.


\subsection{The Auction Setting}\label{subsec:Setting}
The notations used here are similar to the ones used by~\cite{FENG18,DUTTING21,BHARDWAJ23E}. The FC is a single seller intending to sell $m$ homogeneous units of a certain item to $n$ consumers using a forward auction. 

The volume discount bidding is implemented as follows. 
The mechanism solicits volume discount bids from each consumer. The available units are divided into $k$ lots and the consumer places a per-unit bid corresponding to each lot. Thus, if the FC sells more lots, their valuation per unit decreases. In the agricultural domain, this corresponds to savings from the use of bulk transport, warehouse clearance, mass production, etc. 

The consumer $i$'s bid  $b_1^{(i)}$ is applied to all the units if the seller sells at most one lot; i.e. $\ell$ units. For the  units from the second lot,  the bid $b_2^{(i)} $ is used, i.e., a price of $b_1^{(i)}$ per unit for the first $\ell$ units and a price of $b_2^{(i)}$ for the remaining units (up to $\ell$). And so on. 
That is, consumer $i$  submits a volume discount bid in the form of a vector $b^{(i)} = (b^{(i)}_1, b^{(i)}_{2}, \cdots, b^{(i)}_{k})$ of $k$ lots. All the lots are assumed to be of the same size, $\ell := \lfloor \frac{m}{k} \rfloor$ and the extra $m - (k \times l)$ units are assumed to be added to the last lot.
The value of $k$ depends on certain endogenous factors like the packaging method, carton size, nature and price of the items, etc.

Note that this simple model can be used to encapsulate a wide variety of situations. Consider the case where $m = 2500$. 
Say a consumer would like to bid: ((1-500: 20), (501-1500: 18), (1501-2500: 16)) where the first value of the tuple represents the size of that lot and the second value of the tuple represents the bid for that lot. Assume that another consumer would like to procure only $1500$  units, and would like to bid 18 for the first $500$ units and a discount of 1 thereafter. In spite of the difference in their requirements and bidding language, we can incorporate the bids of both these consumers in the following manner. We will have $\ell=500$. The lots would be $[1, 500]$, $[501, 1000]$, $[1001, 1500]$, $[1501, 2000]$, $[2001, 2500]$. The first consumer's bid vector would be $b^{(1)} = (20, 18, 18, 16, 16)$ while that of the second consumer would be $b^{(2)} = (18, 17, 17, -\infty, -\infty)$. If $1800$ units are allocated to the first consumer, then the total cost to this consumer would be $500 \times 20 + 500 \times 18 + 500 \times 18 + 300 \times 16 = 32800$. Similarly, if the allocation for the second consumer is $700$ units, then the corresponding total cost to second consumer would be $500 \times 18 + 200 \times 17 = 12400$.

Given the vector of bids $b = (b^{(1)}, \cdots , b^{(n)})$ as input, the mechanism outputs allocation and payment vectors denoted by the tuple $\langle a(b), p(b) \rangle$. Here,  $a(b)   = (a_1(b), \cdots , a_n(b))$ denotes the allocation vector  and $p(b) = (p_1(b), \cdots , p_n(b))$ denotes the payment vector with each  $a_i(b)$ being the number of units sold to consumer $i$ and $p_i(b)$ being the total payment made by consumer $i$ for the $a_i(b)$ units. 

The consumers have their own private willingness to buy (WTB), which determines the maximum per unit price that the consumer is ready to spend.  For consumer $i$, we denote the WTB by $v^{(i)} = (v^{(i)}_1, ..., v^{(i)}_k)$. Here each $v^{(i)}_j$ represents the valuation or the maximum amount that consumer $i$ is ready to pay for one unit, of the homogeneous items being sold, from lot $j$. Since this is a volume discount auction, the subsequent values of $v_j$ will decrease monotonically, i.e., $v_j \geq v_{j+1} \forall j \in [1,k-1]$. 
These valuations are assumed to be drawn from some prior  distribution $\mathcal{F}$. In our setting, $\mathcal{F}$ is common knowledge among all the consumers and the FC; whereas, the realized vector of valuations $v^{(i)}$ is known privately only to the individual consumer $i$.  
The mechanism is incentive compatible if we have $b^{(i)} = v^{(i)}$. 

The utility of a consumer is defined as a function of their private valuations $v^{(i)}$, allocation $a_i(b)$, and payment $p_i(b)$, and is given by
\begin{equation} \label{eq:cons_util}
    u_i (v^{(i)}; b) =   \sum_{j=1}^{a_i(b)} v^{(i)}_{\lceil j/\ell \rceil} - p_i(b)
\end{equation}

Having defined the utility, we will now provide, for ready reference, the definitions for dominant strategy incentive compatibility (DSIC) and individual rationality (IR). 

A mechanism is DSIC if no agent can gain utility by misrepresenting their valuations, regardless of the strategies adopted by the other agents. That is, \begin{equation}
    u_i (v^{(i)}; (v^{(i)}, b^{(-i)})) \geq u_i (v^{(i)}; b) \qquad \forall v, b, i
\end{equation}

An auction mechanism is called ex-post IR if every agent earns non-negative utility by participating in the auction. We assume that there is no participation cost or auction entry cost. Our proposed neural network architecture is designed to ensure the ex-post IR condition by satisfying
\begin{equation}
    u_i(v^{(i)}; v) \geq 0 \qquad \forall v, i
\end{equation}
It is notable that ex-post IR is the strongest form of individual rationality in mechanism design literature \cite{KRISHNA09}.

Nash social welfare (NSW) of an allocation is defined as the geometric mean of the utilities of the agents under that allocation.
Our goal is to design a mechanism that maximizes NSW, i.e., the product of utilities of all the players are maximized, while ensuring ex-post IR 
and minimum violation of DSIC.  
The Nash Social Welfare has been defined by considering two agents, namely, the FC and a single agent acting as a proxy for the group of bidders. The bidders have been aggregated for the following reason. In any allocation, several of the bidders may end up with zero utility and the NSW will trivially become zero in all such cases. Aggregating all the bidders and working with the sum of the utilities of the bidders avoids this problem while ensuring that the NSW computation also takes the bidders into account. Also note that we are directly maximizing the product of the utilities, rather than the square root of the product, for the sake of convenience. 
\begin{equation}
    nsw = u_{FC} u_{C}
\end{equation}
where $u_{FC}$ and $u_{C}$ are the utility of the FC and the sum of utilities of all the consumers respectively.
This ensures that the mechanism does not unfairly favor either the collective of farmers or the group of consumers. 

The utility of a consumer, $i$ is given in Equation~\ref{eq:cons_util}. Hence, the sum of the utilities of all consumers is 

\begin{equation}
    u_{C} =\sum_{i=1}^n u_i(v^{(i)};b)
\end{equation}

For a particular crop, the reserve price, $p_{res}$, is the lowest unit price at which the FC is willing to sell the crop. The reserve price is announced in advance before the bids are sought. Hence, it acts as a lower bound for the consumers' bids. It may be considered as the FC's per unit valuation for that particular crop. The reserve price may be, for example, the total costing of one unit of the crop with a minimal profit added to it. Hence, assuming that all unsold crop goes waste, the utility of the FC can be calculated as

\begin{equation}
    u_{FC} = \sum_{i=1}^n p_i(b) - p_{res} m 
\end{equation}

Note that the revenue of the FC is different from the utility of the FC. The revenue of the FC is defined as

\begin{equation}
    \text{revenue}_{FC} = \sum_{i=1}^n p_i(b)
\end{equation}

Following section 2.2.2 of~\cite{DUTTING21}, one can guarantee DSIC property by ensuring that the expected ex-post regret for every consumer, $r_i$, is $0$. The expected ex-post regret of the VDA-SAP mechanism is defined as 
\begin{equation}
        r_i = \mathbb{E}_{v \sim  \mathcal{F}}[\max_{b} [ u_i(v^{(i)}; b) - u_i(v^{(i)}; (v^{(i)}, b^{(-i)}))]]
\end{equation}

The regret is computed empirically, which adequately approximates the real regret~\cite{DUTTING21}.

\subsection{Deep Learning Based Formulation}
We propose a neural network based formulation to satisfy individual rationality while minimizing the violation of DSIC, along with some other desirable and practical constraints.  

The goal is to minimize a composite loss function that consists of the following parts; 
the negative of the Nash social welfare, the negative of the FC's revenue, the regret penalty, the envy penalty, the business penalty, and the Lagrangian term (as we use the method of differential multipliers~\cite{PLATT87}) for regret and envy. We also provide models that respectively maximize the revenue of the FC and maximize the revenue of the consumers as the primary goal while minimizing IC violations.

Based on the requirement, different combinations of the various components of the loss function, which are mentioned below, are used. When a particular component is to be maximized, we take its negative value as a part of the loss function that we minimize.
One need not restrict oneself to the usage of the components mentioned in this section. Other nice properties such as egalitarian social welfare can also be included in our methodology by adding similar components to the loss function.
The individual components are each described in further detail through this section. 
\begin{align*}
   &{ \tt revenue_{FC}} = \sum_{i=1}^n p_i(b)\\
  &{ \tt nsw} = u_{FC} \sum_{i=1}^n u_i(v^{(i)};b)\\
    &{ \tt penalty_{regret}} = \rho_{{ \tt regret}} \sum_{i=1}^n \tilde{r}_i^2 \\ 
    &{ \tt penalty_{envy}} = \rho_{{\tt envy}} \sum_{i=1}^n e_i^2\\
\end{align*}
where $penalty_{regret}$ and $penalty_{envy}$ are the regret terms corresponding to regret and envy respectively. 
When trying to minimize envy, the Lagrangian loss used is
\begin{equation}
    { \tt LagrangianLoss } = \sum_{i=1}^n 
    \lambda_{{ \tt regret}}^{(i)} \tilde{r}_i + \lambda_{{\tt envy}}^{(i)} e_i
\end{equation}
and when we are not trying to minimize envy, the Lagrangian loss used is
\begin{equation}\label{eq:LagLossRegret}
    { \tt LagrangianLoss } = \sum_{i=1}^n 
    \lambda_{{ \tt regret}}^{(i)} \tilde{r}_i
\end{equation}
Here, $\tilde{r_i}$ is the empirical regret. We compute $\tilde{r_i}$ by using another optimizer over the bids, coming from the same distribution as $\mathcal{F}$, which maximizes the utility for agent $i$. To approximate the expectation over the distribution $\mathcal{F}$, we maximize the sample mean of regret for the current step.

The equation for the loss function, when only Nash social welfare maximization is considered as an additional objective along with the regret minimization, is 
 \begin{align*}
     &{\tt loss \ = \ -(nsw) \ }  {\tt + \  penalty_{regret} \ } + { \tt LagrangianLoss}
 \end{align*}
where the Lagrangian loss is calculated using Equation~\ref{eq:LagLossRegret}. Other loss terms may be added or removed from this equation as per the requirements.

\subsubsection{Business Constraints}
The seller may wish to impose various business constraints in VDA-SAP. 
For example, the seller may require that at least 3 consumers are allocated at least $20\%$ of the units each. Such a constraint can be introduced by adding a penalty term, while training the network, as shown below. For having a minimum of $s$ consumers, each with an allocation of at least $a_{min}$, the penalty would be
\begin{equation}
    { \tt penalty_{business}} = \rho_{{\tt business}} \sum_{t=1}^{s} \max(0, a_{min}-a^{(t)})
\end{equation}
where $a^{(o)}$ is the $o^{th}$-highest allocation $\forall o \in \{1,2,... s\}$. 
Other business constraints are also possible. For instance, no consumer may be allocated more than $50\%$ of the units. To incorporate various business constraints we add the corresponding penalty for violating those business constraints to the loss function.

\subsubsection{Envy Minimization}
Envy minimization is one of the most popular fairness constraints in auction design.
Envy (or dissatisfaction) for an agent is defined as the maximum utility they could gain if they were given the allocation and payment of some other agent. So the envy for consumer $i$, given the valuation tuple $v = (v^{(1)}, ..., v^{(n)})$ is
\begin{align}
        e_i (v) = \max_{h\in\{1,2,...n\}}& (\sum_{j=1}^{a_h(b)} v^{(i)}_{\lceil j/\ell \rceil} - p_h(b)) - u_i (v^{(i)}; v)
\end{align}
We minimize envy by adding a term for envy in our Lagrangian loss, along with an envy penalty.   

\subsection{Allocation Network and Payment Network}\label{subsec:NetDetails}
\begin{figure}
    \centering
    \includegraphics[width = \columnwidth]{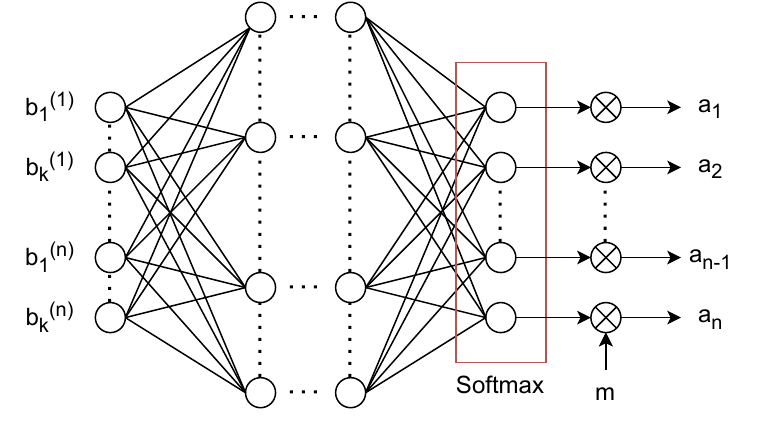}
    \caption{Allocation Network}
    \label{fig:allocationNetwork}
\end{figure}

\begin{figure}
    \centering
    \includegraphics[width = 0.85\columnwidth]{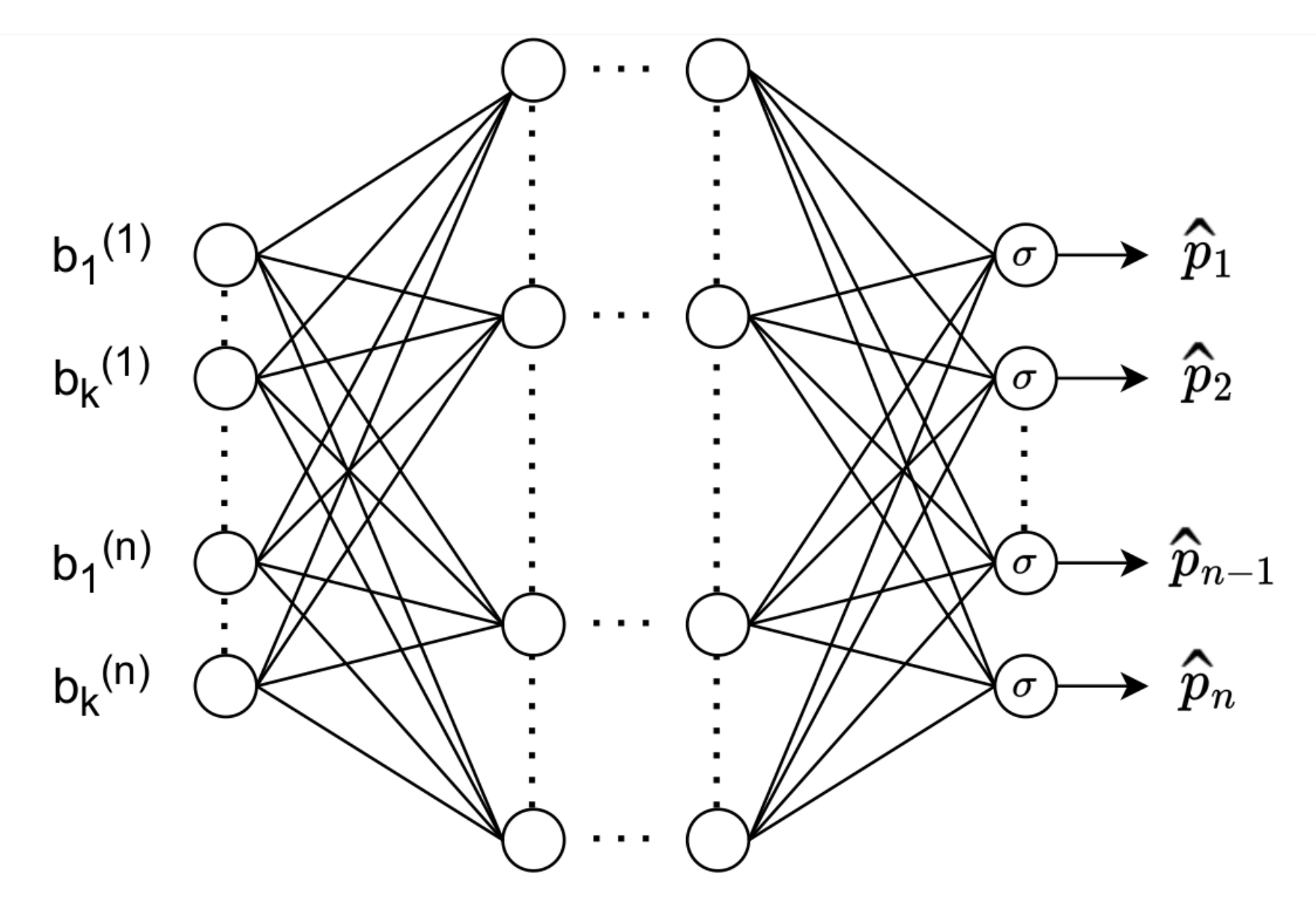}
    \caption{Payment Network}
    \label{fig:paymentNetwork}
\end{figure}

The model consists of two feed-forward networks - an allocation network and a payment network (See Figure~\ref{fig:allocationNetwork} and Figure~\ref{fig:paymentNetwork} for details).  The input for both networks is the $n \times k$ matrix where the $i^{th}$ row is the bid $b^{(i)}$ for consumer $i$.

The output of the allocation network is the allocation tuple described in Section~\ref{subsec:Setting}. The allocation network uses the softmax function to ensure that the allocation tuple is a probability vector. This is multiplied by $m$ to ensure that the allocations across the agents sum up to exactly $m$. If the combined requirements of all the consumers, $CR$ is less than $m$, then the probability vector is multiplied by $CR$ instead of $m$ to ensure that the net allocation is equal to $CR$ and all the consumers are allotted quantities in accordance with their respective requirements.

The output of the payment network is a payment multiplier tuple, $\hat{p} = (\hat{p}_1, ..., \hat{p}_n)$. The amount by which a bid exceeds the reserve price is multiplied by the value in the payment multiplier tuple corresponding to that bid. This value is then added to the reserve price to get the corresponding payment. Hence, the total payment made by consumer $i$ would be 
\begin{equation}\label{eq:paymentTuple}
p_i (b) = p_{res} e + \hat{p}_i (b)\sum_{j=1}^{a_i(b)} (b^{(i)}_{\lceil j/\ell \rceil} - p_{res})
\end{equation}
where $e = (1, ..., 1)$ is an n-tuple where all elements have the value $1$. 
Each $\hat{p}_i$ is guaranteed to be within the range $[0,1]$ in order to ensure IR for the consumers. This is ensured by adding a sigmoid layer at the end of the payment network. IR for the farmers is ensured by keeping the payments higher than the reserve price in Equation~\ref{eq:paymentTuple}.

\subsection{Training Procedure}
In all our experiments $5$ layers, with $80$ to $100$ neurons in each layer, were used for both the payment and allocation networks. The Adam optimizer was used for training the network weights and stochastic gradient descent was used to learn the Lagrangian parameters.

During the training phase, we performed nested optimizations. For one step of optimization over the network weights, we executed $R$ steps of optimization over the bids to compute the empirical regret. Here, we used gradient ascent to maximize the value of empirical regret that a misreporting consumer would experience. This empirical regret computation and the use of Lagrangian parameters is the same as what is proposed by D\"{u}tting, Feng, Narasimhan, Parkes, and Ravindranath~\cite{DUTTING21}.

\subsection{Why a Deep Learning Approach?}
NSW maximization is widely known for many nice properties~\cite{CARAGIANNIS19} but finding an outcome that maximizes NSW is NP-hard~\cite{RAMEZANI09}. The methodology we follow essentially transforms a mechanism design problem into an optimization problem. But even an approximate solution to this optimization problem cannot be found as the constraints that we deal with, such as business constraints and envy minimization, are nonlinear and, unfortunately, linear approximations do not work well with such constraints. 
Moreover, the linear approximation will have an exponential number of variables. 

The deep learning technique that we have invoked here does involve a substantial effort of training. However, it has the advantage that a single substantial effort of training  will amortize the computational complexity over a large number of VDA-SAP experiments that may be required. 

\section{Experimental Results}\label{sec:Exp}
In this section, we describe our experimentation with an FC-mediated market for a perishable vegetable such as tomato, brinjal, chili pepper, or for seasonal fruits. As described in Section~\ref{sec:Prosper}, the FC gathers details about the produce to be sold from a number of registered farmers and invites volume discount bids from potential consumers.

For our experimentation, we considered $1000$ units of the produce, for example, $1000$ kg of chili pepper. 
This is an example of a typical volume of a given produce that a farmer might sell.
All prices are in  US \$. 
We consider five consumers – they could be major consumers like retail chains or medium level stores like community grocery stores. 
We determine the valuation of a single unit of the produce for each consumer to be drawn from the uniform distribution U$[3.5,4.5]$. The consumers place volume discount bids, which have been formed to cover all representative scenarios.
\begin{itemize}
\item Consumer 1  specifies a flat per unit price (uniform price) for the entire range of $[1,1000]$ units and does not bid with any discount. 
\item Consumer 2 bids with a discount of 5\% if the purchase volume is in the range of $[501,1000]$ units. This discount is over the base price that applies for the purchase volume range $[1,500]$. 
\item Consumer 3  bids with a discount of 3\% for any volume in the range of $[301,600]$ units and 6\% for a volume in the range of $[601,1000]$. The discount is over the base price that applies for the volume range $[1,300]$. 
\item Consumer 4 bids with a discount of 2\%, 4\% and 6\% for the volume ranges $[251,500], [501,750],$ and $[751,1000]$  units, respectively. This discount is over the base price that applies for the purchase volume range $[1,250]$. 
\item Consumer 5 bids with a discount of 2\%, 4\%, 6\% and 8\% for the volume ranges $[201,400], [401,600], [601,800],$ and $[801,1000]$ units, respectively. This discount is over the base price that applies for the purchase volume range $[1,200]$. 
\end{itemize}

An important parameter to be selected is the reserve price. A lower reserve price hurts the farmer while a higher reserve price hurts the consumers. We have logically chosen \$3 per unit after observing the results for different values of reserve prices. 
Refer to Table~\ref{tab:results} for numerical results. The following performance metrics are computed: 
\begin{itemize}
\item FC revenue, which is the total revenue of the FC and is also the total payment made by  all the consumers.
\item FC utility, which is the amount by which FC revenue 
exceeds the sum total of reserve prices for the produce sold.
\item Total consumer utility, which is the sum of the utilities of all consumers. A consumer's utility is the amount by which the consumer's valuation of all the units allotted to the consumer exceeds the payment made for those units.
\item Social welfare, which is the sum of the utility of the FC and utilities of all the consumers.
\item Nash social welfare, which is the product of the FC's utility and the sum of utilities of all the consumers. NSW is generally defined as the geometric mean of the player utilities. However, for the sake of convenience, 
we have not calculated the square root and are directly reporting the product of the utilities.
\item Regret, which is the maximum possible fractional (per unit) increase in utility, of any consumer, achieved by  making a non-truthful bid.
\item Envy, which is the maximum possible fractional (per unit) increase in utility of any consumer if they were to be allotted another consumer's allocation and payment.
\end{itemize} 

Each column in Table~\ref{tab:results} holds the values for one of the six metrics mentioned above. These values were calculated after averaging the values obtained for over 6000 runs where, in each run, the valuations of the FC and the consumers were drawn from the uniform distribution U$[3.5,4.5]$.  The five rows of the table correspond to five different auctions that are relevant in this scenario: 
VCG  Auction; FC  Optimal Auction; Consumer Optimal Auction; NSW Maximizing Auction; and NSW Maximizing \& Envy Minimizing Auction. The first three auctions were included to serve as baseline auctions for making performance comparisons. While the VCG auction can be implemented analytically, the rest of the auctions were implemented using the deep learning approach described in Section~\ref{sec:DLApproach}.

\begin{table*}[t]
    \centering
    \caption{Various utility measures (in US \$) on sale of 1000 units using different auction mechanisms}
    \label{tab:results}
    \begin{tabular}{cccccccc}\toprule
         & & \textbf{Total} & & \textbf{Nash} & & & \\
         & \textbf{FC} & \textbf{Consumer} & \textbf{Social} & \textbf{Social} & \textbf{FC}  & & \\
         & \textbf{Utility} & \textbf{Utility} & \textbf{Welfare} & \textbf{Welfare} & \textbf{Revenue} & \textbf{Regret} & \textbf{Envy} \\ \midrule 
         \textbf{VCG} & 763 & 264 & 1027 & 193016 & 3763 & 0.0 & 0.1817 \\ \hline
         \textbf{FC Optimal} & 859 & 96 & 955 & 79292 & 3859 & 0.0499 & 0.2274 \\ \hline
         \textbf{Consumer Optimal} & 1 & 875 & 876 & 243 & 3001 & 0.0001 & 0.1985 \\ \hline
         \textbf{NSW Maximizing} & 397 & 603 & 1000 & 243444 & 3397 & 0.0291 & 0.1378 \\ \hline
         \textbf{NSW Maximizing \& } & & & & & & & \\
         \textbf{Envy Minimizing} & 251 & 748 & 999 & 190990 & 3251 & 0.0265 & 0.0129 \\ \bottomrule
    \end{tabular}
\end{table*}

\subsection{Baseline Auctions}

\subsubsection*{VCG Auction}
Our first baseline auction is the standard VCG mechanism, which maximizes social welfare  while satisfying DSIC and IR. The results show that the utility of the FC is much higher than the combined utility of the consumers, thus the mechanism appears to be more FC friendly (that is farmer friendly) and the consumers may not be excited by this auction mechanism.

\subsubsection*{FC Optimal Auction}
Our second baseline auction maximizes the expected revenue of the FC subject to satisfying IC and IR, via loss function minimization. The following loss function is used
 \begin{align*}
     {\tt loss \ = \ -(revenue_{FC}) \ } & {\tt + \  penalty_{regret} \ } \\
     & + { \tt LagrangianLoss}
 \end{align*}
This auction produces the highest possible revenue for the FC (subject to the stated constraints) and the farmers will be delighted with such an auction (see Table~\ref{tab:results}). 
At the same time, such an auction may turn away consumers from participating in the auction. 

\subsubsection*{Consumer Optimal Auction}
Our third baseline auction maximizes the revenue of the consumers subject to satisfying IC and IR, via loss function minimization. The following loss function is used
 \begin{align*}
     &{\tt loss \ = \ revenue_{FC} \ }  {\tt + \  penalty_{regret} \ } + { \tt LagrangianLoss}
 \end{align*}

This auction produces the highest possible utility for the consumers (see Table~\ref{tab:results}). 
However, such an auction will prove to be unattractive to the farmers since the FC's revenue is affected. 

\subsection{NSW Maximizing Auction}
We now consider an auction that maximizes NSW, ensures IC as well as IR, and satisfies certain business constraints. The loss function to be minimized for such an auction would be 
 \begin{align*}
     {\tt loss \ = \ -(nsw) \ } & {\tt + \  penalty_{regret} \ } + { \tt penalty_{business} \ } \\ 
     &+ { \tt LagrangianLoss}
 \end{align*} 
The business constraints used in our experiments try to ensure that the number of consumers who are allotted the produce remains between a minimum number and a maximum number. 
It should be noted that other business constraints can also be used equally effectively by modifying the $penalty_{business}$ term in accordance with the required business constraints.
The motivation for having a constraint on minimum number of consumers is to spread the business among competing consumers to avoid favoring only one or two aggressive consumers, providing equal opportunities to small and big consumers. The constraint on maximum number of consumers is to optimize logistics costs. 

Let us call this auction NSW maximizing auction.
We find from Table~\ref{tab:results} that the social welfare of the VCG auction is 1027 while that of the NSW maximizing auction is 1000; however, there is a perceptible difference in the utilities of the FC and consumers between these two auctions. Recall that in the case of the VCG auction, the utilities were loaded in favor of the FC. In the case of the NSW maximizing auction, the utilities are 397 for the FC and 603 for the consumers which is more balanced than the values of 763 and 264, respectively, in the case of the VCG auction. The NSW for the NSW maximizing auction is 243,444 which is clearly superior to 193,016 of the VCG auction. The FC revenue is 3763 in the case of the VCG auction while it is 3397 in the case of the NSW maximizing auction. The decreased revenue of the FC is compensated by increased utility of consumers. The envy of the NSW maximizing auction is 0.1378 compared to 0.1817 of the VCG auction.  To summarise, the NSW maximizing auction achieves a better balance in the utilities for the FC and consumers, achieves better NSW, results in more utility for the consumers, and leads to less envy, at the cost of some decrease in the revenue of the FC. It is clear that the NSW maximizing auction will be more acceptable to both farmers and consumers than the VCG auction as the VCG auction empirically shows a certain  bias towards the farmers in our experiments.

\subsubsection*{NSW Maximizing \& Envy Minimizing Auction}  
Though NSW balances social welfare maximization with fairness of allocation, in many situations, fairness may have to be accorded a high priority. For instance, in agricultural markets, there is a critical need to ensure that nobody goes out of business due to aggressive bidding by powerful players. Envyfreeness is a way of ensuring this. 
The loss function to be minimized for such an auction would be
 \begin{align*}
     {\tt loss \ = \ -(nsw) \ } & {\tt + \  penalty_{regret} + penalty_{envy} \ } \\
     & + { \tt penalty_{business} \ }
      {\tt + LagrangianLoss}
 \end{align*}
The last row in Table~\ref{tab:results} shows the results for this new auction. As expected, this auction achieves a significantly small value of envy, however, at the cost of reduced utility, reduced revenue for the FC, and reduced value of NSW. The consumer utility is higher than that of the NSW maximizing auction. In situations where fairness may have to be accorded a high priority, this auction is preferable.

\section{Conclusion and Future Work}\label{sec:Conclusion}
In this paper, we have designed a versatile mechanism for selling harvested produce of farmers to potential consumers through intermediation by farmer collectives using volume discount auctions. 
The designed auctions maximize Nash social welfare subject to IC, IR, fairness, and business constraints.
Detailed experimentation on these auctions show the efficacy of the mechanisms designed. 
Our work provides clear evidence that the proposed mechanisms will be more attractive than existing traditional methods. They bring many benefits such as ensuring farmer welfare, consumer delight, inducing honesty in bidding, utilizing scale economies, selecting deserving consumers, and the possibility to ensure fairness of allocation.

Other than volume discounts, agricultural markets commonly offer combinatorial discounts. Here, discounts may be requested by consumers or offered by FCs based on specific combinations of items sold. These discounts may correspond to savings in transport and packaging for the FC, or may help the FC clear stocks of particular slow-moving items. 
A deep learning based framework for combinatorial auctions is an important future direction. 
Integrating combinatorial discounts and volume discounts would also pose an interesting challenge and would be an apt representation for some real world agricultural market situations.

While the motivation behind our proposed method is based on several conversations with farmers and visits to FCs, we note that further exploration of real world dynamics that might impact the auction design would need extensive randomized controlled trials involving farmers and FCs. This will also include conducting detailed surveys to understand user experiences. A detailed look at the on-ground implementation of such mechanisms is a major direction for future work, which will help remove impediments and realize the potential benefits of adopting the proposed mechanisms.

\balance

This paper has also not addressed the issue of explainability of the deep learning model used for the design of VDA-SAP. This is another important direction for future work. 

\bibliographystyle{ACM-Reference-Format} 
\bibliography{AGRIBIB}

@BOOK{NARAHARI14,
  author = {Y. Narahari},
  title = {Game Theory and Mechanism Design},
  publisher = {IISc Press (Bengaluru, India) and The World Scientific (Singapore)},
  year = {2014}
}

@ARTICLE{GALE62,
  author = {D. Gale and L. S. Shapley},
  title = {College Admissions and the Stability of Marriage},
  journal = {The American Mathematical Monthly},
  volume = {69},
  number = {1},
  year = {1962},
  pages = {9--15},
}

@book{KRISHNA09,
  title={Auction theory},
  author={Krishna, Vijay},
  year={2009},
  publisher={Academic press}
}

@article{BHARDWAJ23E,
  title={Designing Fair, Cost-optimal Auctions based on Deep Learning for Procuring Agricultural Inputs through Farmer Collectives},
  author={Bhardwaj, Mayank Ratan and Ahmed, Bazil and Diwakar, Prathik and Ghalme, Ganesh and Narahari, Y},
  journal={2023 IEEE 19th International Conference on Automation Science and Engineering (CASE)},
  pages={1--8},
  year={2023},
  organization={IEEE}
}

@ARTICLE{MILGROM89,
  author = {P.~Milgrom},
  title = {Auctions and bidding: A primer},
  journal = {Journal of Economic Perspectives},
  volume = {3}, number ={3},
  year = {1989},
  pages = {3--22}
}

@inproceedings{PLATT87,
  title={Constrained differential optimization},
  author={Platt, John and Barr, Alan},
  booktitle={Neural Information Processing Systems},
  pages={612-621},
  year={1987}
}

@online{IDINSIGHT19,
    author = {Nair, Divya and Singh, Rupika and Jalote, Sumedha and Sharma, Vinod Kumar and Thompson, Will},
    title = {Enabling Farmer Producer Companies’ Success in Marketing},
    url = {https://www.idinsight.org/publication/enabling-farmer-producer-companies-success-in-marketing/},
    publisher = {IDInsight},
    date = {10},
    month = {11},
    year = {2020}
}

@inproceedings{DEVI15,
  title={E-mandi implementation based on gale-shapely algorithm for perishable goods supply chain},
  author={Devi, S Prasanna and Narahari, Y and Viswanadham, Nukala and Kiran, S Vinu and Manivannan, S},
  booktitle={2015 IEEE International Conference on Automation Science and Engineering (CASE)},
  pages={1421--1426},
  year={2015},
  organization={IEEE}
}

@article{LEVI22,
  title={Improving Farmers’ Income on Online Agri-platforms: Evidence from the Field},
  author={Levi, Retsef and Rajan, Manoj and Singhvi, Somya and Zheng, Yanchong},
  journal={Available at SSRN 3486623},
  year={2022}
}

@misc{ECONOMIST17,
  title={Ethiopia’s state-of-the-art commodity exchange},
  author={The Economist},
  year={2017}
}

@article{STRZKEBICKI15,
  title={The electronic marketplace as the element of the agricultural market infrastructure},
  author={Strzkebicki, Dariusz},
  journal={Problems of Agricultural Economics},
  number={1\_2015},
  year={2015}
}

@INPROCEEDINGS{NEWMAN18,
  title={Designing and evolving an electronic agricultural marketplace in Uganda},
  author={Newman, Neil and Bergquist, Lauren Falcao and Immorlica, Nicole and Leyton-Brown, Kevin and Lucier, Brendan and McIntosh, Craig and Quinn, John and Ssekibuule, Richard},
  booktitle={Proceedings of the 1st ACM SIGCAS Conference on Computing and Sustainable Societies},
  pages={1--11},
  year={2018}
}

@online{ENAM,
    author = " e NAM ",
    url = "https://www.enam.gov.in/web/"
}

@inproceedings{RAMEZANI09,
  title={Nash social welfare in multiagent resource allocation},
  author={Ramezani, Sara and Endriss, Ulle},
  booktitle={International Workshop on Agent-Mediated Electronic Commerce},
  pages={117--131},
  year={2009},
  organization={Springer}
}

@article{DUTTING21,
  title={Optimal Auctions through Deep Learning},
  author={Paul Dütting and Zhe Feng and Harikrishna Narasimhan and David C. Parkes and Sai S. Ravindranath},
  journal={Communications of the ACM}, VOLUME=64, Number=8, PAGES ={109--116},
  year={2021}
}

@INPROCEEDINGS{DUTTING19,
  title={Optimal Auctions through Deep Learning},
  author={Paul Dütting and Zhe Feng and Harikrishna Narasimhan and David C. Parkes and Sai S. Ravindranath},
  BOOKTITLE={Proceedings of the 36th International Conference on Machine Learning}, 
  PAGES={1706--1715},
  YEAR={2019}
}

@INPROCEEDINGS{DUTTING15,
  title={Payment rules through discriminant-based classifiers},
  author={D{\"u}tting, Paul and Fischer, Felix and Jirapinyo, Pichayut and Lai, John K and Lubin, Benjamin and Parkes, David C},
  booktitle={ACM Transactions on Economics and Computation},
  volume={3},
  number={1},
  article={5},
  pages={5:1--5:41},
  year={2015},
  publisher={ACM New York, NY, USA}
}

@inproceedings{FENG18,
  title={Deep learning for revenue-optimal auctions with budgets},
  author={Feng, Zhe and Narasimhan, Harikrishna and Parkes, David C},
  booktitle={Proceedings of the International Conference on Autonomous Agents and Multiagent Systems (AAMAS 2018)},
  pages={354--362},
  year={2018}
}

@article{YUEN23,
  title={Extending the characterization of maximum Nash welfare},
  author={Yuen, Sheung Man and Suksompong, Warut},
  journal={Economics Letters},
  volume={224},
  pages={111030},
  year={2023},
  publisher={Elsevier}
}

@article{CARAGIANNIS19,
  title={{The unreasonable fairness of maximum Nash welfare}},
  author={Caragiannis, Ioannis and Kurokawa, David and Moulin, Herv{\'e} and Procaccia, Ariel D and Shah, Nisarg and Wang, Junxing},
  journal={ACM Transactions on Economics and Computation (TEAC)},
  volume={7},
  number={3},
  pages={1--32},
  year={2019},
  publisher={ACM New York, NY, USA}
}

@TECHREPORT{ZHANG21,
Author={Zhanhao Zhang},
Title={A Survey of Online Auction Mechanism Design Using
Deep Learning Approaches},
Institution={arXiv Preprint},
Number={arXiv:2110.06880v1},
Year={2021}}

@article{SANKAR24,
  title={Deep Learning Meets Mechanism Design: Key Results and Some Novel Applications},
  author={Sankar, V Udaya and Rao, Vishisht Srihari and Narahari, Y},
  journal={arXiv preprint arXiv:2401.05683},
  year={2024}
}

@TECHREPORT{EarthObservingSystem22,
Author={Earth Observing System},
Title={Agricultural Cooperatives: Specifics, Role, Pros, and Cons},
Institution={https://eos.com/blog/agricultural-cooperatives/},
Year={2022}}

@inproceedings{VISWANADHAM12,
  title={{Mandi electronic exchange: Orchestrating Indian agricultural markets for maximizing social welfare}},
  author={Viswanadham, Nukala and Chidananda, Sridhar and Narahari, Yadati and Dayama, Pankaj},
  booktitle={2012 IEEE International Conference on Automation Science and Engineering (CASE)},
  pages={992--997},
  year={2012},
  organization={IEEE}
}

@misc{JADHAV19, 
    title={Just 14\% of farmers registered on eNAM platform}, 
    url={https://www.thehindubusinessline.com/economy/agri-business/just-14-of-farmers-registered-on-enam-platform/article28363454.ece}, 
    journal={The Hindu BusinessLine}, 
    author={Radheshyam Jadhav}, 
    year={2019}, 
    month={July},
    date={10}
}

@article{MAINA23,
  title={Access to and utilization of local digital marketing platforms in potato marketing in Kenya},
  author={Maina, Florence and Mburu, John and Nyang'anga, Hillary},
  journal={Heliyon},
  volume={9},
  number={8},
  year={2023},
  publisher={Elsevier}
}

\end{document}